\documentclass[aps,two column]{revtex4}%
\usepackage{amsfonts}
\usepackage{amsmath}
\usepackage{amssymb}
\usepackage{graphicx}%
\providecommand{\U}[1]{\protect\rule{.1in}{.1in}}
\begin{document}
\title{Surface morphology and magnetic anisotropy of Fe/MgO(001) films deposited at
oblique incidence}
\author{Qing-feng Zhan and Chris Van Haesendonck}
\affiliation{Laboratorium voor Vaste-Stoffysica en Magnetisme, Katholieke Universiteit
Leuven, Celestijnenlaan 200\,D, BE-3001 Leuven, Belgium}
\author{Stijn Vandezande and Kristiaan Temst}
\affiliation{Instituut voor Kern- en Stralingsfysica, Katholieke Universiteit Leuven,
Celestijnenlaan 200\thinspace D, BE-3001 Leuven, Belgium}

\begin{abstract}
We have studied surface morphology and magnetic properties of Fe/MgO(001)
films deposited at an angle varying between 0$^{\text{o}}$ and 60$^{\text{o}}$
with respect to the surface normal and with azimuth along the Fe[010] or the
Fe[110] direction. Due to shadowing, elongated grains appear on the film
surface for deposition at sufficiently large angle. X-ray reflectivity reveals
that, depending on the azimuthal direction, films become either rougher or
smoother for oblique deposition. For deposition along Fe[010] the pronounced
uniaxial magnetic anisotropy (UMA) results in the occurrence of
\textquotedblleft reversed\textquotedblright\ two-step and of three-step
hysteresis loops. For deposition along Fe[110] the growth-induced UMA is much
weaker, causing a small rotation of the easy axes.

\end{abstract}
\maketitle

Magnetic anisotropy of epitaxial films and its relationship to surface
morphology have attracted much attention in recent years~\cite{RPP-59-1409}.
Both film properties are intimately related to the molecular beam epitaxy
(MBE) deposition process. Oblique incidence deposition results via a
self-shadowing effect~\cite{PRL-98-046103, PRB-77-235423} in the formation of
grains in the plane of the film that are elongated perpendicular to the
incident flux direction and with aspect ratio increasing at larger deposition
angle with respect to the surface normal~\cite{EPL-75-119}. Consequently, an
in-plane uniaxial magnetic anisotropy (UMA) with easy axis perpendicular to
the incident flux direction is induced during growth of the magnetic
films~\cite{EPL-75-119, APL-77-2030, APL-87-082505}. UMA was found to play an
important role in determining the magnetization reversal in thin films of
cubic systems~\cite{AM-19-323, JAP-78-7210}. Depending on the strength and
orientation of the UMA, hysteresis curves with one, two and three steps are
observed in various films at different field orientation. The appearance of
the steps can be explained in terms of nucleation and propagation of domain
walls (DWs)~\cite{JAP-78-7210, PRL-79-4018}. Consequently, understanding the
influence of oblique incidence growth is very important because of its ability
to control both magnetic anisotropy and surface
morphology~\cite{APL-91-092502, PRB-77-115427, APL-87-073105}.

Although shadowing effects have been studied in many magnetic systems,
including Fe on MgO(001)~\cite{APL-66-2140} and Co on
Cu(001)~\cite{PRB-63-104431}, the plane of oblique incidence was always kept
parallel to the in-plane cubic easy axes of the magnetic layers. Here, we
report on the influence of oblique deposition on the surface morphology and
magnetic properties of Fe/MgO(001) films for deposition azimuth both along the
Fe[010] and Fe[110] in-plane directions. Deposition along [010] turns out to
be more effective in producing elongated grains and introducing UMA than
deposition along [110]. A considerable growth-induced UMA along [010] induces
the occurrence of \textquotedblleft reversed\textquotedblright\ two-step and
of three-step hysteresis loops.

Fe (001) oriented films were grown on MgO(001) substrates in an MBE system
with base pressure below $1\times$10$^{-10}\,\mathrm{mbar}$. The substrates
were annealed for one hour at $700\,^{\text{o}}\mathrm{C}$ and held at
$150\,^{\text{o}}\mathrm{C}$ during deposition. The films were deposited using
an e-beam gun at a rate of 0.1~\AA /s, as monitored by a quartz crystal
oscillator. The incident Fe beam was at an angle varying between $0^{\text{o}%
}$ to $60^{\text{o}}$ with respect to the surface normal and with azimuth
either along the Fe[010] or the Fe[110] direction. In Figs.~1(a) and 1(b) we
present the surface topography measured \textit{in situ} by scanning tunneling
microscopy (STM) for Fe films grown at an incidence angle of $30^{\text{o}}$
and with azimuth along [010] and [110], respectively. The film surface
consists of approximately square grains about $15\,\mathrm{nm}$ in size. The
grain edges have a preferred in-plane orientation along $\left\langle
010\right\rangle $. When the deposition angle is increased to $60^{\text{o}}$,
Fe surface grains for deposition along [010] are obviously elongated
perpendicular to the incident flux direction with typical length of
$15\,\mathrm{nm}$ and typical width of $3\,\mathrm{nm}$ (see Fig.~1(c)). The
elongated grains are believed to result from a redistribution of the incident
flux due long-range attractive forces~\cite{PRL-82-4038, PRB-67-165425}.
During evaporation the incident Fe atoms arrive preferentially on top of
already formed grains rather than behind these grains. This shadowing effect
is also observed for film deposition at $60^{\text{o}}$ with azimuth along
[110]. The Fe grains on the surface are rhombic in shape with typical length
of $15\,\mathrm{nm}$ and typical width of $6\,\mathrm{nm}$ (see Fig.~1(d)).
Based on the STM images we conclude that the self-shadowing effect can be
neglected for a deposition angle below $30^{\text{o}}$. Moreover, the
self-shadowing effect is anisotropic, i.e., depends on the azimuth of the
incident flux direction. The surface grains for growth along [010] have a
larger aspect ratio than the grains for deposition along [110].

The anisotropy of the shadowing effect is confirmed by \textit{ex situ} X-ray
reflectivity (XRR). Before removing the Fe/MgO(001) layers from the vacuum
chamber, the layers are capped with a $4\,\mathrm{nm}$ thick protective Au
layer. In Fig.~2(a) we present the X-ray data collected on a PANalytical
X'Pert Pro diffractometer. When increasing the deposition angle from
0$^{\text{o}}$ to 60$^{\text{o}}$, the oscillating X-ray intensity is rapidly
damped for the samples grown along [010], while the oscillations are slightly
enhanced for the samples grown along [110]. By fitting the X-ray curves, we
quantitatively obtain the root-mean-square (rms) roughness for both the lower
interface between the Fe layer and the MgO substrate (see Fig.~2(b)) and the
upper interface between the Fe layer and the Au layer (not shown). The
roughness of both interfaces rapidly increases when increasing the deposition
angle for the samples deposited along [010], but slightly decreases for the
samples deposited along [110]. Thus, the shadowing effect is able to not only
roughen but also smoothen Fe/MgO(001) films, depending on the deposition
geometry. Although we used the same evaporation rate and the same nominal
thickness ($15\,\mathrm{nm}$) when growing the Fe layers, the actual thickness
obtained from the X-ray simulations decreases from $14\,\mathrm{nm}$ to
$8\,\mathrm{nm}$ when the deposition angle increases from $0^{\text{o}}$ to
$60^{\text{o}}$ (see Fig.~2(b)).

Due to the elongated shape of the Fe grains in the Fe/MgO(001) films an
additional UMA is superimposed upon the intrinsic cubic anisotropy $K_{1}$ of
Fe. Depending on the azimuthal angle, the growth-induced UMA may have a
component $K_{u1}$ along [010], a component $K_{u2}$ along [110], or both.
When $K_{u1}<<K_{1}$ and $K_{u2}<K_{1}$, the component $K_{u2}$ rotates the
position of the overall easy axes backwards with respect to the uniaxial hard
axis over an angle $\delta$ that is given by $\delta=\frac{1}{2}\sin
^{-1}(K_{u2}/K_{1})$~\cite{APL-91-122510}, as illustrated in Fig.~2(c). The
magnetic properties of the Fe/MgO(001) films were characterized \textit{ex
situ} by the longitudinal and transverse magneto-optical Kerr effect (MOKE)
for different field orientation $\phi$ as defined in Fig.~2(c).

For the category of samples grown with deposition angle $\leq30^{\text{o}}$
and along [010] and for the samples with deposition angle $\leq49^{\text{o}}$
and along [110], one-step and two-step hysteresis loops are measured at
different $\phi$. For the typical loops presented in Figs.~3(a) and 3(b),
which are obtained at $\phi=15^{\text{o}}$ and $125^{\text{o}}$ for the film
deposited at normal incidence, the switching routes that occur for increasing
field are [$\overset{-}{1}$00]$\rightarrow$[100] and [0$\overset{-}{1}%
$0]$\rightarrow$[$\overset{-}{1}$00]$\rightarrow$[010], respectively. The
corresponding spin orientations are given by the arrows that are enclosed in a
square in Fig.~3. For this category of samples, the switching fields for
one-step ($H_{c1}$) and two-step ($H_{c1}$, $H_{c2}$) loops as a function of
$\phi$ are all similar to the fields of the film deposited at normal
incidence, which are presented in Fig.~4(a). From the symmetries of the
coercive fields and the angles where one-step and two-step loops are observed,
we infer that a weak uniaxial anisotropy is present along [010].

For the two samples deposited at an angle exceeding $49^{\text{o}}$ and along
[010], we observe, apart from one step loops, \textquotedblleft
reversed\textquotedblright\ two-step, and three-step loops as well. The
typical loops presented in Figs.~3(c) and 3(d) are obtained at $\phi
=95^{\text{o}}$ and $65^{\text{o}}$, respectively, for the sample deposited at
$60^{\text{o}}$ and along [010]. The switching routes for increasing field are
[0$\overset{-}{1}$0]$\rightarrow$[100]$\rightarrow$[010] and [0$\overset{-}%
{1}$0]$\rightarrow$[$\overset{-}{1}$00]$\rightarrow$[100]$\rightarrow$[010],
respectively. Obviously, the magnetization $M$ for the \textquotedblleft
reversed\textquotedblright\ two-step loops rotates in the opposite direction
(i.e., the switching route changes from clockwise to counter-clockwise or vice
versa) when compared to the normal two-step loops observed in the same half
quadrant of $\phi$. The $\phi$ dependence of the experimental switching fields
of one-step ($H_{c1}$), \textquotedblleft reversed\textquotedblright\ two-step
($H_{c3}$, $H_{c4}$), and three-step ($H_{c3}$, $H_{c4}$, $H_{c5}$) loops is
symmetric about $\left\langle 010\right\rangle $ for the two samples, as
illustrated in Fig.~4(b) for the sample grown at $60^{\text{o}}$ and along
[010]. The symmetries of the coercive fields and the angles where different
loops occur, suggest that the extra UMA is oriented along [010].

For the sample deposited at $60^{\text{o}}$ and along [110], only one-step and
two-step loops are observed at different $\phi$. The easy axes are found to
slightly deviate from the cubic easy axes by $\delta=3^{\text{o}}$. The
magnetization does not switch exactly over $90^{\text{o}}$, but over
$90^{\text{o}}\pm2\delta$. We link the presence of a component $K_{u2} \neq0$
to the oblique deposition along [110]. The influence of $K_{u2} \neq0$ also
clearly emerges in the $\phi$ dependence of the experimental switching fields
($H_{c1}$, $H_{c2}$) presented in Fig.~4(c), which is symmetric about
$\left\langle 110\right\rangle $. However, the range of angles $\phi$ where
one-step loops occur, is not symmetric about either [110] or [010], which
indicates that $K_{u1}$ cannot be neglected.

Recently, we have shown that both the $180^{\text{o}}$ and the $90^{\text{o}%
}\pm2\delta$ magnetic transitions, which occur in the hysteresis loops with
different number of steps, are mediated by successive or separate
$90^{\text{o}}\pm2\delta$ DW nucleations~\cite{Unpublished}. The UMA component
$K_{u1}$ and the DW nucleation energies $\epsilon_{90^{\text{o}}\pm2\delta}$
can be evaluated by fitting the $\phi$ dependence of the experimental
switching fields~\cite{JAP-78-7210, APL-91-122510, Unpublished}. In Figs.~4(a)
to 4(c) we present the fitting curves for three typical samples. The fitting
parameters for all our Fe/MgO(001) films are plotted in Fig.~4(d). For films
with $K_{u2}=0$, i.e., $\delta=0$, $\epsilon_{90^{\text{o}}}/M$ remains
constant at about $0.35\,\mathrm{mT}$ for a deposition angle $\leq
30^{\text{o}}$ (along [010] as well as along [110]), but increases to
$0.79\,\mathrm{mT}$ for the deposition at $60^{\text{o}}$ and along [010].
This indicates that the DW nucleation energy is sensitive to the Fe layer
thickness, but not the deposition geometry. The small enhancement of
$\epsilon_{90^{\text{o}}}$ can be accounted for by the thickness reduction of
Fe layers deposited at larger angle. For the sample grown at $60^{\text{o}}$
and along [110], the DW nucleation energy can take the two values
$\epsilon_{90^{\text{o}}\pm2\delta}$ due to the shift of the easy axes
resulting from $K_{u2}\neq0$. For the set of samples deposited with azimuthal
angle along [110], $K_{u1}/M$ is smaller than $0.1\,\mathrm{mT}$. Such weak
UMA cannot be avoided even for films deposited at normal incidence. We
therefore assume that this small $K_{u1}$ results from the growth imposed by
the MgO(001) substrate and not from the oblique deposition. The influence of
the shadowing effect on $K_{u2}$ is also small. Only for the sample deposited
at $60^{\text{o}}$ and along [110], a small $K_{u2}$, which is about
$0.1\,K_{1}$, is observed experimentally. Changes in deposition geometry are
clearly more effective for deposition along [010]. $K_{u1}/M$ slowly increases
from $0.07\,\mathrm{mT}$ to $0.21\,\mathrm{mT}$ when increasing the deposition
angle from $0^{\text{o}}$ to $30^{\text{o}}$, and then quickly rises to
$2.70\,\mathrm{mT}$ at $49^{\text{o}}$ and $7.45\,\mathrm{mT}$ at
$60^{\text{o}}$. The stronger UMA obviously results from changes in the
surface morphology, where clearly elongated grains start to be formed at a
deposition angle between $30^{\text{o}}$ and $49^{\text{o}}$ for the growth
along [010]. It is more difficult to impose an elongated grain structure for
oblique deposition along [110] because of the preferential directions [100]
and [010] for the growth of Fe grains on MgO(001).

For different $\phi$ the switching fields are predicted to be proportional to
$\left\vert K_{u1}\pm\epsilon_{90^{\text{o}}}\right\vert $~\cite{JAP-78-7210,
PRL-79-4018}. Moreover, the field orientation where different kinds of
hysteresis loops can be observed, is determined by the condition $\tan
\phi<K_{u1}/\epsilon_{90^{\text{o}}}$, where $0^{\text{o}}<\phi<45^{\text{o}}$
for the one-step loops and $45^{\text{o}}<\phi<90^{\text{o}}$ for the
three-step loops, respectively~\cite{Unpublished}. Consequently, for samples
deposited along [010] the coercive fields also rapidly increase due to the
enhancement of $K_{u1}$ resulting from the oblique deposition. On the other
hand, the range of $\phi$ for the occurrence of one-step and three-step loops
becomes wider as predicted by theory. Since the observation of the
\textquotedblleft reversed\textquotedblright\ two-step and of the three-step
switching events requires $K_{u1}>\epsilon_{90^{\text{o}}}$, we conclude that
the critical angle for deposition of Fe/MgO(001) films for which we can
observe these two kinds of loops is between $30^{\text{o}}$ and $49^{\text{o}%
}$ for deposition along [010]. For deposition along Fe[110], the
growth-induced UMA is much weaker, causing a small rotation of the easy axes.

\noindent{\Large \textbf{FIGURE CAPTIONS}} \newline\newline\noindent FIG.~1.
(Color online) STM images for Fe films deposited (a) at $30^{\text{o}}$ and
along [010], (b) at $30^{\text{o}}$ and along [110], (c) at $60^{\text{o}}$
and along [010], and (d) at $60^{\text{o}}$ and along [110]. The size of the
images is $200\,\mathrm{nm}\times200\,\mathrm{nm}$ for (a) and (b), and
$100\,\mathrm{nm}\times100\,\mathrm{nm}$ for (c) and (d). \newline%
\newline\noindent FIG.~2. (Color online) (a) XRR spectra for films deposited
at $0^{\text{o}}$ (black), $11^{\text{o}}$ (red), $30^{\text{o}}$ (green),
$49^{\text{o}}$ (blue), and $60^{\text{o}}$ (magenta). The azimuthal angle for
deposition is parallel to [010] (lower curves) and [110] (upper curves),
respectively. (b) The rms roughness and the thickness for Fe layers deposited
along [010] (squares) and along [110] (dots), as inferred from simulations of
the XRR spectra. (c) Definition of the angles that are used to describe a film
with in-plane cubic anisotropy and an additional UMA. \newline\newline%
\noindent FIG.~3. (Color online) Longitudinal ($\parallel$) and transverse
($\perp$) MOKE loops with (a) one step and (b) two steps, obtained at
$\phi=15^{\text{o}}$ and $125^{\text{o}}$ for the film deposited at normal
incidence, and (c) \textquotedblleft reverse\textquotedblright\ two-step and
(d) three-step loops obtained at $\phi=95^{\text{o}}$ and $65^{\text{o}}$ for
the sample deposited at $60^{\text{o}}$ and along [010]. The blue (red) curves
are for applied fields varying from negative (positive) to positive (negative)
saturation. The orientation of the Fe spins in the switching processes is
represented by the arrows enclosed in a square. \newline\newline\noindent
FIG.~4. (Color online) The experimental switching fields $H_{c1}$
({\Large $\circ$}), $H_{c2}$ ({\small $\square$}), $H_{c3}$
({\large $\triangledown$}), $H_{c4}$ ({\Large $\diamond$}), and $H_{c5}$
({\large $\vartriangle$}) as a function of $\phi$, and the corresponding
theoretical curves for the samples deposited (a) at normal incidence, (b) at
60$^{\text{o}}$ and along [010], and (c) at 60$^{\text{o}}$ and along [110].
(d) The fitting parameters $K_{u1}$ ({\Large $\circ$}), $\epsilon
_{90^{\text{o}}}$ ({\small $\square$}), $\epsilon_{90^{\text{o}}-2\delta}$
({\Large $\triangleleft$}), and $\epsilon_{90^{\text{o}}+2\delta}$
({\Large $\triangleright$}) for samples grown along [010] (upper panel), and
along [110] (lower panel).

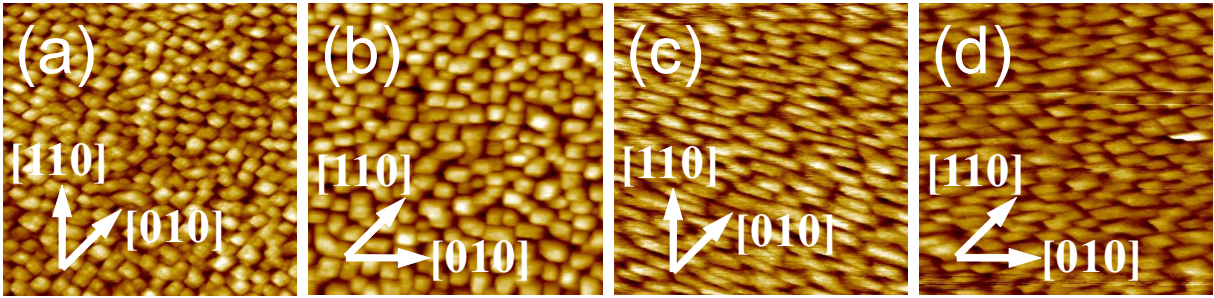
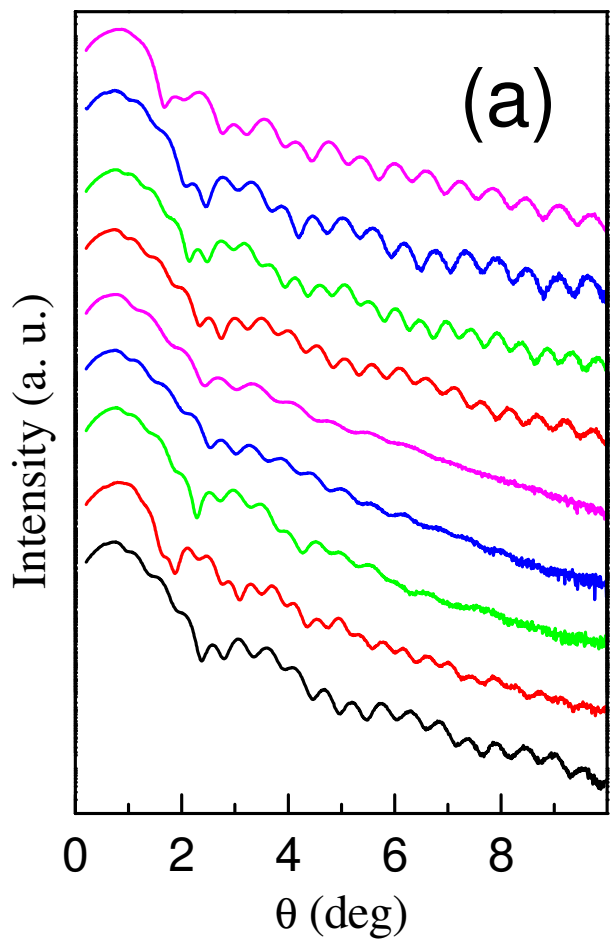
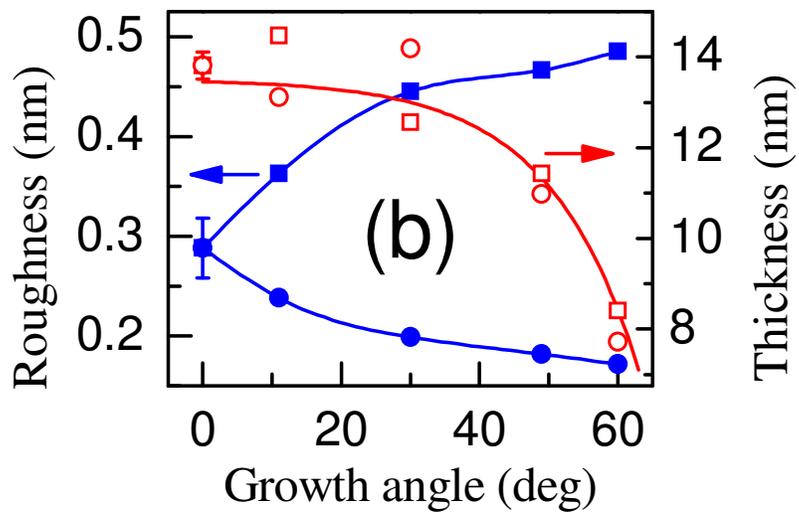
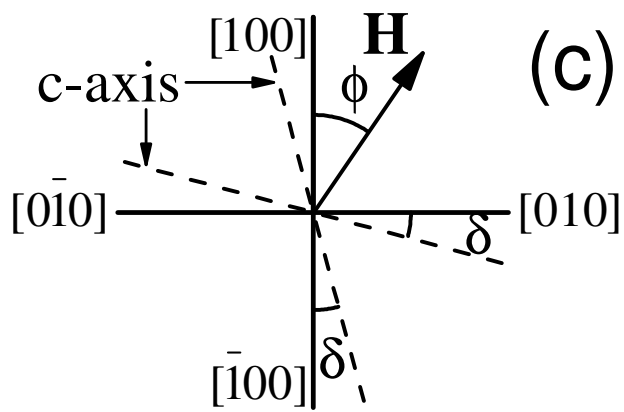
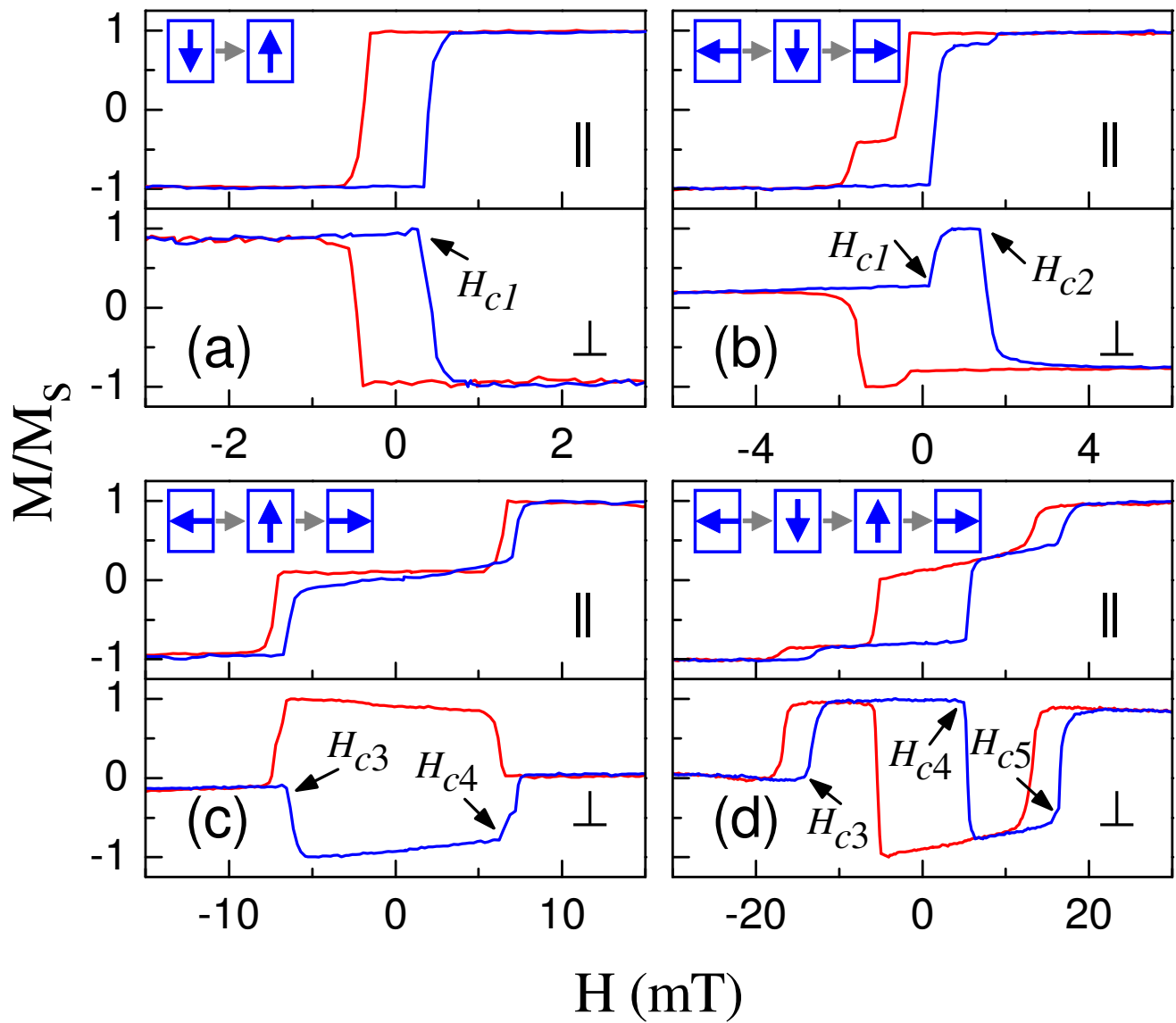
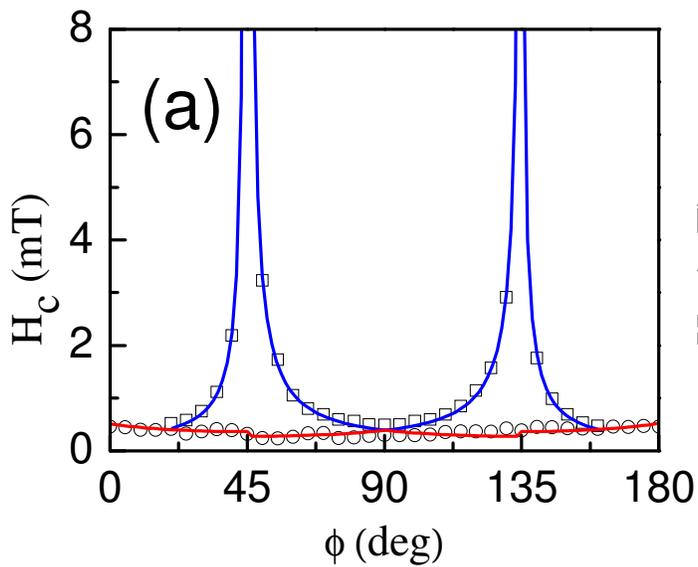
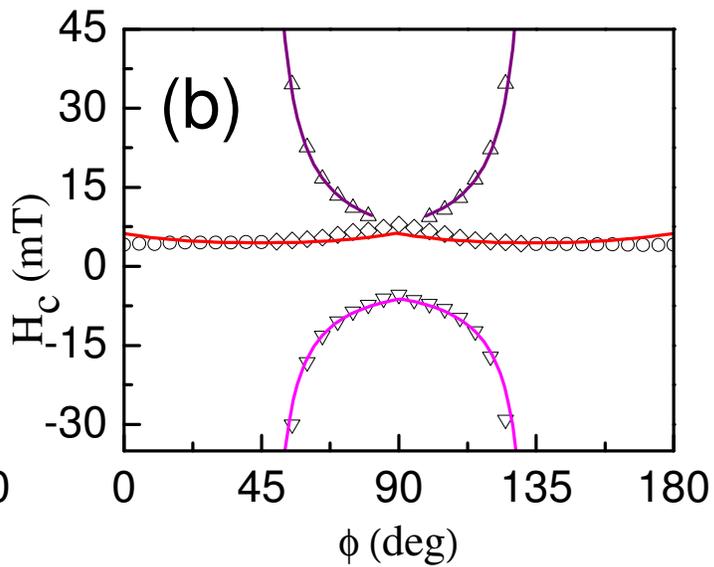
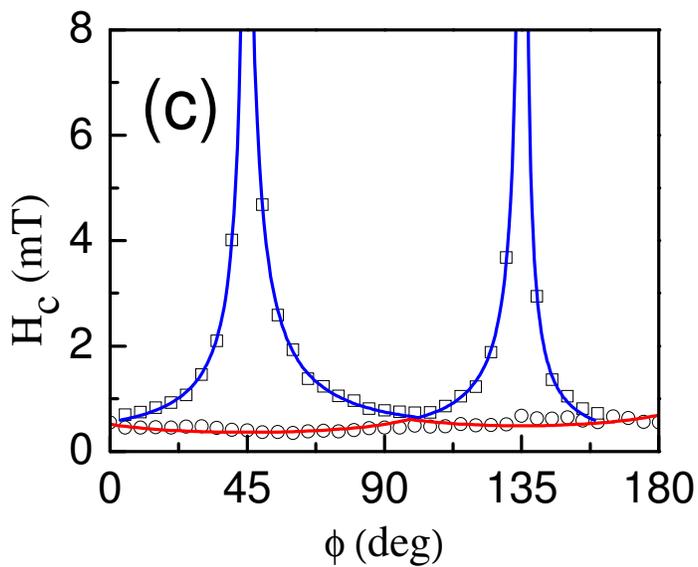
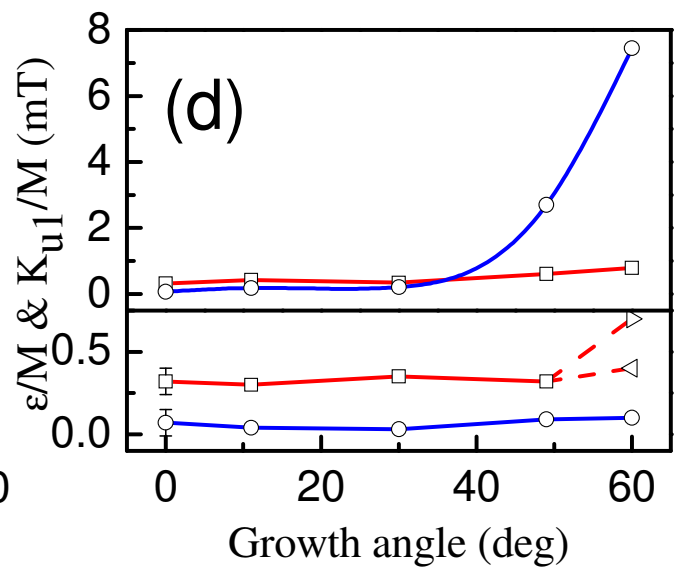
\end{document}